\documentclass[twocolumn,english,prd,showpacs,showkeys,preprintnumbers,notitlepage]{revtex4}
\usepackage[T1]{fontenc}
\usepackage[latin9]{inputenc}
\usepackage{amsmath}
\usepackage{graphicx}
\usepackage{amssymb}
\usepackage{esint}

\makeatletter
\@ifundefined{textcolor}{}
{%
 \definecolor{BLACK}{gray}{0}
 \definecolor{WHITE}{gray}{1}
 \definecolor{RED}{rgb}{1,0,0}
 \definecolor{GREEN}{rgb}{0,1,0}
 \definecolor{BLUE}{rgb}{0,0,1}
 \definecolor{CYAN}{cmyk}{1,0,0,0}
 \definecolor{MAGENTA}{cmyk}{0,1,0,0}
 \definecolor{YELLOW}{cmyk}{0,0,1,0}
 }

\usepackage{xcolor}\PassOptionsToPackage{normalem}{ulem}\usepackage{ulem}

\makeatletter

\usepackage{xcolor}\usepackage{babel}

\usepackage[unicode=true, pdfusetitle,
 bookmarks=true,bookmarksnumbered=false,bookmarksopen=false,
 breaklinks=false,pdfborder={0 0 1},backref=false,colorlinks=false]{hyperref}

\def\GeV{\,\mbox{GeV}}\newcommand{\beq}{\begin{equation}}\newcommand{\eeq}{\end{equation}}\newcommand{\beqn}{\begin{eqnarray}}\newcommand{\eeqn}{\end{eqnarray}}

\makeatother

\usepackage{babel}

\begin{document}

\title{Diffractive neutrino-production of pions on nuclei:\\
 Adler relation within the color-dipole description}

\author{B.Z.~Kopeliovich}

\email{Boris.Kopeliovich@usm.cl}

\author{Iv\'an~Schmidt}

\email{Ivan.Schmidt@usm.cl}

\author{M.~Siddikov}

\email{Marat.Siddikov@usm.cl}

\affiliation{Departamento de F\'{i}sica, Centro de Estudios Subat\'omicos, y Centro
Cient\'{i}fico - Tecnol\'ogico de Valpara\'{i}so, Universidad T\'ecnica
Federico Santa Mar\'{i}a, Casilla 110-V, Valpara\'{i}so, Chile}

\preprint{USM-TH-296}


\begin{abstract}
Effects of coherence in neutrino-production of pions off nuclei are
studied employing the color dipole representation and path integral
technique. If the nucleus remains intact, the process is controlled
by the interplay of two length scales. One is related to the pion
mass and is quite long (at low $Q^{2}$), while the other, associated
with heavy axial-vector states, is much shorter. The Adler relation
is found to be broken at all energies, but especially strongly at
$\nu\gtrsim10\GeV$, where the cross section is suppressed by a factor
$\sim A^{-1/3}$. On the contrary, in a process where the recoil nucleus
breaks up into fragments, the Adler relation turns out to be strongly
broken at low energies, where the cross section is enhanced by a factor
$\sim A^{1/3}$, but has a reasonable accuracy at higher energies,
where all the coherence length scales become long. 
\end{abstract}
\maketitle

\section{Introduction }

Due to the $V-A$ structure of weak interactions high-energy neutrinos
serve as a source of the axial current. Unfortunately, because of
the smallness of the neutrino-hadron cross-sections experimental data
have been quite scarce until recently, mostly being limited to the
total cross-sections. With the launch of the new high-statistics experiments
like MINER$\nu$A at Fermilab~\cite{Drakoulakos:2004gn}, now the
neutrino-hadron interactions may be studied with a better precision.

The properties of the vector current have been well studied, mostly
in collisions of charged leptons with protons and nuclei in processes
of deep inelastic scattering (DIS), deeply virtual Compton scattering
(DVCS), real Compton scattering (RCS) and meson production. The structure
of the axial current is less known.

According to Adler relation (AR) \cite{Adler:1964yx,Adler:1966gc},
the cross section of neutrino interaction at zero virtuality is proportional
to the cross-section of pion interaction on the same target and with
the same final hadronic state, \begin{equation}
\left.\frac{d\sigma_{\nu p\to lF}}{d\nu\, dQ^{2}}\right|_{Q^{2}=0}=\frac{G_{F}^{2}}{2\pi^{2}}\, f_{\pi}^{2}\,\frac{E_{\nu}-\nu}{E_{\nu}\nu}\,\sigma_{\pi p\to F}(\nu),\label{eq:Adler}\end{equation}
 where $G_{F}=1.166\times10^{-5}\GeV^{-2}$ is the electro-weak Fermi
coupling; $F$ denotes the final hadronic state; $E_{\nu}$ and $\nu$
are the energy of the neutrino and transferred energy in the target
rest frame, respectively.

Nuclear effects in diffractive neutrino-production of pions, coherent
($\nu A\to\pi A$) and incoherent ($\nu A\to\pi A^{*}$), were calculated
in \cite{belkov} based on the AR and Glauber eikonal approximation.
The results were different from previous calculations performed in
\cite{Rein:1982pf}, which contradicted the AR and which were based
on an incorrect model for nuclear effects (see critical discussion
in \cite{belkov}).

Recently, a large deviation from the AR predictions for coherent and
incoherent diffractive neutrino-production of pions at high energies
was discovered in \cite{Kopeliovich:2011rk} using a simple two-channel
toy model which contains only axial meson and pion. This deviation
is caused by initial and final state interactions, called absorptive
corrections, which are very strong for large rapidity gap processes
like diffractive production. The onset of these corrections is controlled
by the coherence length $l_{c}=2\nu/m_{a}^{2}$, where $\nu$ is the
transferred energy (or the pion energy), and $m_{a}\sim1\GeV$. The
AR is at work only if the coherence length is short compared to the
nuclear size, $l_{c}\ll R_{A}$, i.e. at low energies. At higher energies
the value of the cross section considerably drops, by a factor $\sim A^{-1/3}$,
compared to the AR prediction. In this paper we extend the result
obtained in~\cite{Kopeliovich:2011rk} and demonstrate that it is
valid in a realistic color dipole model.

The nuclear shadowing effect for the total neutrino-nucleus interaction
at low $Q^{2}$ was first calculated in \cite{bell} within a specific
optical model, which was essentially oversimplified. It was also calculated
within the Glauber-Gribov theory \cite{glauber,Gribov:1968jf} in
\cite{k-shad,gransasso,shad-jetp}, and in this case it gave good
agreement with data from the WA59 experiment \cite{wa59}. This calculation
was based on the AR, which in this case has no absorptive corrections,
and is expected to be rather accurate.

In this paper we describe the neutrino-nucleus interactions within
the color dipole approach which was proposed in~\cite{Kopeliovich:1981pz}
for description of the high-energy scattering processes. The dipole
representation is especially simple and effective at high energies,
where the dipole separation does not fluctuate during propagation
through the nucleus, being ''frozen'' by Lorentz time dilation.
Besides, at high energies, or small Bjorken $x$ in deep-inelastic
scattering (DIS), gluonic exchanges with the target dominate in the
scattering amplitude. The phenomenological dipole cross section is
usually fitted to HERA data for the proton structure function at small
$x$, and it is risky to use at at lower energies, where Reggeons,
i.e. quark-antiquark exchanges become important, and should be explicitly
added to the dipole cross section. Eventually, at energies as low
as $\sqrt{s}\lesssim2$~GeV the model is not valid and one should
refer to other models which contain explicit contributions of resonances
(see e.g.~\cite{Lalakulich:2006sw,Lalakulich:2006yn} and a recent
review in~\cite{Nakamura:2011rt}). We perform a more rigorous calculation
than was done in \cite{Kopeliovich:2011rk}, where the different coherence
lengths were introduced by hand.

The color dipole approach was tested in a number of photon-nucleon
and photon-nucleus processes, such as Deep Inelastic Scattering~\cite{GolecBiernat:1998js,GolecBiernat:2004xu},
Drell-Yan reaction \cite{raufeisen} heavy meson production~\cite{Hufner:2000jb},
as well as deeply virtual Compton scattering (DVCS), real Compton
scattering (RCS), double deeply virtual Compton scattering (DDVCS)
on the nucleons and nuclei (See e.g. \cite{Kopeliovich:2008ct,Kopeliovich:2009cx,Kopeliovich:2010xm,Kopeliovich:2010sa,Machado:2008tp,Machado:2008zv,Machado:2009cd}),
giving a reasonable description of the total and differential cross-sections.
Also, the color dipole model has been applied to the description of
the neutrino physics in~\cite{Fiore:2005bp,Fiore:2005yi,Fiore:2008cc,Fiore:2008nj,GayDucati:2008hi,GayDucati:2008zzc,Ducati:2006vh,Machado:2007wq}.
Single-pion production by neutrinos on a proton target has been studied
recently within the color dipole approach in~\cite{Kopeliovich:2011xw}.

Here we extend the results obtained for neutrino-proton interactions
in~\cite{Kopeliovich:2011xw} to the nuclei using the Glauber-Gribov
approach~\cite{glauber,Gribov:1968jf}. While in the high-energy
({}``frozen'') regime the shadowing corrections are given by the
trivial exponential attenuation factor, we use an approach which is
also valid at intermediate energies, where the dipole size fluctuates
during propagation through the nucleus. As was discussed earlier,
we do not consider the region of very low energies ($\sqrt{s}\lesssim2$~GeV)
due to limitations of the model and absence of explicit $s$-channel
resonances in the model.

The paper is organized as follows. In Section~\ref{sec:proton} for
the sake of completeness we give the formulas which are used for evaluation
of the color dipole amplitudes on the proton. In Section~\ref{sec:NuclearEffects}
we discuss the framework which was used for evaluation of nuclear
corrections. In Section~\ref{sec:Results} we present results and
draw conclusions.

\section{Diffractive pion production on a proton}

\label{sec:proton}

In this section we present a brief survey of the formulas for evaluation
of the neutrino cross-section on a proton target. More details can
be found in~\cite{Kopeliovich:2011rk}. The pion production cross-section
in the neutrino-proton collisions has the form \begin{equation}
\frac{d^{3}\sigma_{\nu p\to l\pi p}}{dtdQ^{2}dx_{{\rm Bj}}}=\frac{G_{F}^{2}xy^{2}}{32\pi^{3}Q^{4}}\frac{L_{\mu\nu}\left(W_{\mu}^{A\to\pi}\right)^{*}W_{\nu}^{A\to\pi}}{\left(1-\frac{q^{2}}{M_{W}^{2}}\right)^{2}\sqrt{1+\frac{4m^{2}x_{{\rm Bj}}^{2}}{Q^{2}}}},\label{eq:disgma_XSection}\end{equation}
 where $L_{\mu\nu}$ is the axial lepton tensor, and $W_{\nu}^{A\to\pi}$
is the amplitude of pion production by axial current on the proton
target. In the color dipole model this amplitude has the form \begin{eqnarray}
W_{\mu}^{A\to\pi}\left(s,\Delta,Q^{2}\right) & = & \left(g_{\mu\nu}-\frac{q_{\mu}q_{\nu}}{q^{2}-m_{\pi}^{2}}\right)\int\limits _{0}^{1}d\beta_{1}d\beta_{2}\label{eq:W_amplitude_definition}\\
 & \times & \int d^{2}r_{1}d^{2}r_{2}\bar{\Psi}_{f}^{\pi}\left(\beta_{2},\vec{r}_{2}\right)\nonumber \\
 & \times & \mathcal{A}^{d}\left(\beta_{1},\vec{r}_{1};\beta_{2},\vec{r}_{2};\Delta\right)\Psi_{\nu}^{i}\left(\beta_{1},\vec{r}_{1}\right)\nonumber \end{eqnarray}
 where $\bar{\Psi}_{f}^{\pi}$ and $\Psi_{\nu}^{i}$ are the distribution
amplitudes of the pion and axial current respectively; $\Delta$ is
the 4-momentum transfer in the dipole-proton scattering, , and $\mathcal{A}^{d}(...)$
is the dipole scattering amplitude.

The distribution amplitudes are essentially nonperturbative objects.
We parametrize them in the form derived in~\cite{Dorokhov:2006qm,Anikin:2000rq,Dorokhov:2003kf,Goeke:2007j,Kopeliovich:2011rv}.
The dipole scattering amplitude $\mathcal{A}^{d}(...)$ in~(\ref{eq:W_amplitude_definition})
is a universal object, which depends only on the target, but not on
the projectile and final states. In addition to the axial current
contribution, in~(\ref{eq:W_amplitude_definition}) the contribution
of the vector current should be also present. This contribution involves
a poorly known helicity flip dipole amplitude $\tilde{\mathcal{A}}_{d}$,
which is small~\cite{Buttimore:1998rj} anyway, and therefore we
neglect it\textbf{.} Moreover, at small $Q^{2}$ the vector current
contribution is suppressed by a factor that goes as $Q^{2}$.

At high energies in the small angle approximation, $\Delta/\sqrt{s}\ll1$,
the quark separation and fractional momenta $\beta$ are preserved,
so \begin{eqnarray}
\mathcal{A}^{d}\left(\beta_{1},\vec{r}_{1};\beta_{2},\vec{r}_{2};Q^{2},\Delta\right) & \approx & \delta\left(\beta_{1}-\beta_{2}\right)\delta\left(\vec{r}_{1}-\vec{r}_{2}\right)\label{eq:DVCSIm-BK}\\
 & \times & \left(\epsilon+i\right)\Im m\, f_{\bar{q}q}^{N}\left(\vec{r},\vec{\Delta},\beta,s\right)\nonumber \end{eqnarray}
 where $\epsilon$ is the ratio of the real to imaginary parts, and
for the imaginary part of the elastic dipole amplitude we employ the
model developed in \cite{Kopeliovich:2007fv,Kopeliovich:2008nx,Kopeliovich:2008da,Kopeliovich:2008ct},\begin{widetext}
\begin{eqnarray}
\Im m\, f_{\bar{q}q}^{N}\left(\vec{r},\vec{\Delta},\beta,s\right)=\frac{\sigma_{0}(s)}{4}\exp\left[-\left(\frac{B(s)}{2}+\frac{R_{0}^{2}(s)}{16}\right)\vec{\Delta}_{\perp}^{2}\right]\left(e^{-i\beta\vec{r}\cdot\vec{\Delta}}+e^{i(1-\beta)\vec{r}\cdot\vec{\Delta}}-2e^{i\left(\frac{1}{2}-\beta\right)\vec{r}\cdot\vec{\Delta}}e^{-\frac{r^{2}}{R_{0}^{2}(s)}}\right).\label{eq:ImF}\end{eqnarray}
 \end{widetext} and the phenomenological functions $\sigma_{0}(s),\, R_{0}^{2}(s),\, B(s)$
are fitted to DIS, real photoproduction and $\pi p$ scattering data.

In the forward limit, $\Delta\to0$, the imaginary part of the amplitude
(\ref{eq:DVCSIm-BK}) reduces to the saturated form \cite{GolecBiernat:1998js}
of the dipole cross section, \begin{eqnarray}
\sigma_{d}(r,s) & = & \Im m\, f_{\bar{q}q}^{N}\left(\vec{r},\vec{\Delta},\beta,s\right)\label{eq:Sigma_QQ_GPD}\\
 & = & \sigma_{0}(s)\left(1-\exp\left(-\frac{r^{2}}{R_{0}^{2}(s)}\right)\right).\nonumber \end{eqnarray}

The calculation of the differential cross section also involves the
real part of scattering amplitude, which according to \cite{Bronzan:1974jh}
is related to the imaginary part as \begin{equation}
\mathcal{R}e\,{f(\Delta=0)}=s^{\alpha}\tan\left[\frac{\pi}{2}\left(\alpha-1+\frac{\partial}{\partial\ln s}\right)\right]\frac{\Im m\,{f(\Delta=0)}}{s^{\alpha}}.\label{eq:BronzanFul}\end{equation}
 In the model under consideration the imaginary part of the forward
dipole amplitude indeed has a power dependence on energy, $\mathcal{I}m\, f(\Delta=0;\, s)\sim s^{\alpha}$,
so (\ref{eq:BronzanFul}) simplifies to \begin{eqnarray}
\frac{\mathcal{R}e\,\mathcal{A}}{\Im m\,\mathcal{A}} & =\tan\left(\frac{\pi}{2}(\alpha-1)\right)\equiv\epsilon.\end{eqnarray}

This fixes the phase of the forward scattering amplitude, which we
retain for nonzero momentum transfers, assuming similar dependences
for the real and imaginary parts.

\section{Nuclear effects}

\label{sec:NuclearEffects}

\subsection{Quark shadowing}

\label{sec:Quark shadowing}

Nuclear shadowing in hard reactions originates mainly from the contribution
of soft interactions (if any). In the color dipole model, the soft
contribution arises from the so called aligned jet configurations
\cite{bjorken}, corresponding to $\bar{q}q$ fluctuations very asymmetric
in sharing the photon momentum, $\beta\ll1$. Such fluctuations, having
large transverse separation, are the source of quark shadowing \cite{k-povh}.
They are suppresses for longitudinally polarized currents, and do
not exist if the hard scale is imposed by the heavy quark mass rather
than by virtuality $Q^{2}$. This clearly shows that quark shadowing
is a higher twist effect. The leading twist shadowing effects arise
from the higher Fock components containing gluons, $|\bar{q}qg\rangle$.
Indeed, the gluon carries a small fraction of the total momentum,
therefore the mean transverse separation of the $\bar{q}q$ and gluon
is large even at high $Q^{2}$. We provide more details on gluon shadowing
in the next section.

As was discussed in~\cite{Kopeliovich:2011rk}, neutrino-production
of pions on nuclei is controlled by two characteristic length scales,
the coherence length for pion production, \begin{equation}
l_{c}^{\pi}=\frac{2\nu}{Q^{2}+m_{\pi}^{2}},\label{eq:l_pi}\end{equation}
 and the coherence length related to excitation in the intermediate
state of axial-vector states, like the $a_{1}$-meson, or a $\rho\pi$
pair. The latter has an invariant mass distribution, which peaks close
to the $a_{1}$ mass, and can be treated as an effective $a$-pole
\cite{asymmetry,deck,belkov,marage,Kopeliovich:2011rk}, \begin{equation}
l_{c}^{a}=\frac{2\nu}{Q^{2}+m_{a}^{2}}.\label{eq:l_a}\end{equation}
 For large virtuality $Q^{2}\gg m_{a}^{2}$ the nuclear effects depend
on one coherence length $l_{c}^{\pi}\approx l_{c}^{a}$. However,
for $m_{\pi}^{2}\lesssim Q^{2}\ll m_{a}^{2}$ we have $l_{c}^{\pi}\gg l_{c}^{a},$
so there are three different regimes for the coherence effects. 
\begin{itemize}
\item For $l_{c}^{a}\ll l_{c}^{\pi}\ll R_{A}$ the coherence length is small
and there is no shadowing. The cross-section of coherent pion production
(the nucleus remains intact) vanishes, and the incoherent cross section
(the nucleus decays to fragments) is a simple sum of the cross-sections
on separate nucleons. 
\item In the intermediate regime of $l_{c}^{\pi}\gg R_{A}$, $l_{c}^{a}\ll R_{A}$
the $\bar{q}q$ dipole is produced instantaneously inside the nucleus
and then evolves into the pion wave function. For the distribution
amplitude of the dipole, we may use the distribution amplitude evaluated
in the IVM. 
\item If $l_{c}^{\pi}\gg l_{c}^{a}\gg R_{A}$ the axial current fluctuates
into a $\bar{q}q$ dipole long before the production of the pion,
and this meson may scatter on the nucleons. In this regime one can
treat the dipole size as {}``frozen'' by Lorentz time dilation,
what considerably simplifies the calculations. As was discussed in
detail in~\cite{Kopeliovich:2011rk}, the Adler theorem~(\ref{eq:Adler})
is severely broken in this regime, even for $Q^{2}=0$, due to large
absorptive corrections. 
\end{itemize}
The theoretical description of the transition region, where the lifetime
of a $\bar{q}q$ fluctuation cannot be either neglected or considered
to be sufficiently long to apply the {}``frozen'' size approximation,
is the most difficult task. In this regime a $\bar{q}q$ dipole propagates
through the nuclear medium with a varying size. In this paper we employ
the description of shadowing developed in~\cite{Kopeliovich:1998gv}
and based on the light-cone Green function technique \cite{kz91}.
The propagation of a color dipole in the nuclear medium is described
as a motion in an absorptive potential, \begin{eqnarray}
\left[i\frac{\partial}{\partial z_{2}}\right.\!\! & + & \!\!\left.\frac{\Delta_{\perp}\left(r_{2}\right)-\varepsilon^{2}}{2\nu\alpha\left(1-\alpha\right)}+U(r_{2},z_{2})\right]G\left(z_{2},\vec{r}_{2};z_{1},\vec{r}_{1}\right)\nonumber \\
 & = & i\delta\left(z_{2}-z_{1}\right)\delta^{\left(2\right)}\left(\vec{r}_{2}-\vec{r}_{1}\right).\label{eq:W-definition}\end{eqnarray}
 where the Green function $G\left(z_{2},\vec{r}_{2};z_{1},\vec{r}_{1}\right)$
describes the probability amplitude for the propagation of dipole
state with size $r_{1}$ at the light-cone starting point $z_{1}$
to the dipole state with size $r_{2}$ at the light-cone point $z_{2}$;
$\epsilon^{2}=\alpha(1-\alpha)Q^{2}+m_{q}^{2}$, and the imaginary
part of the light-cone potential describes absorption in the nuclear
medium, \begin{equation}
{\Im m}\, U(r,z)=-\frac{1}{2}\,\sigma_{\bar{q}q}(r)\,\rho_{A}(b,z).\label{4.270}\end{equation}

In this paper we assume that the real part of the scattering potential
is zero. This approximation is justified for large $Q^{2}$, and for
small $Q^{2}$ the real part should be added, as was done in~\cite{kst2}.

Then the shadowing correction to the amplitude of the coherent pion
production gets the form \begin{equation}
\mathcal{A}\left(\Delta\right)=\int dz\, d^{2}b\,\rho_{A}(b,z)e^{ib\cdot\Delta}\left(F_{1}\left(b,z\right)-F_{2}\left(b,z\right)\right),\label{eq:correction}\end{equation}
 where

\begin{eqnarray*}
F_{1}\left(b,\, z\right) & = & \int d\alpha d^{2}r_{1}d^{2}r_{2}\bar{\Psi}_{f}\left(\alpha,r_{2}\right)G\left(+\infty,\vec{r}_{2};z,\vec{r}_{1}\right)\\
 & \times & \sigma_{\bar{q}q}\left(r_{1}\right)\Psi_{i}\left(\alpha,r_{1}\right)\\
F_{2}\left(b,\, z\right) & = & \int^{z}dz_{2}d\alpha d^{2}r_{1}d^{2}r_{2}d^{2}r_{3}\bar{\Psi}_{f}\left(\alpha,r_{3}\right)\\
 & \times & G\left(+\infty,\vec{r}_{3};z,\vec{r}_{2}\right)\sigma_{\bar{q}q}\left(r_{2}\right)G\left(z,\vec{r}_{2};z_{2},\vec{r}_{1}\right)\\
 & \times & \rho_{A}\left(z_{2},b\right)\sigma_{\bar{q}q}\left(r_{1}\right)\Psi_{i}\left(\alpha,r_{1}\right)\end{eqnarray*}

Equation~(\ref{eq:W-definition}) is quite complicated and in the
general case can be solved only numerically \cite{nemchik}. However
in some cases an analytic solution is possible. For example, in the
limit of a long coherence length, $l_{c}\gg R_{A}$, relevant for
high-energy region, one can neglect the {}``kinetic'' term~$\propto\Delta_{r_{2}}G\left(z_{2},r_{2};z_{1},r_{1}\right)$
in~(\ref{eq:W-definition}), and the Green function formalism reproduces
the well-known eikonal formula in the {}``frozen'' approximation
\cite{kz91}: \begin{eqnarray}
G\left(z_{2},r_{2};z_{1},r_{1}\right) & = & \delta^{2}\left(\vec{r}_{2}-\vec{r}_{1}\right)\label{eq:W-frozen}\\
 & \times & \exp\left(-\frac{1}{2}\sigma_{\bar{q}q}\left(r_{1}\right)\int\limits _{z_{1}}^{z_{2}}d\zeta\,\rho_{A}\left(\zeta,b\right)\right)\nonumber \end{eqnarray}
 so the amplitude~(\ref{eq:correction}) simplifies to

\begin{eqnarray}
\mathcal{A}(s,\Delta_{\perp}) & = & 2\int d^{2}b\, e^{i\vec{\Delta}_{\perp}\cdot\vec{b}_{\perp}}\int\limits _{0}^{1}d\alpha\, d^{2}r\,\bar{\Psi}_{f}\left(\alpha,r\right)\Psi_{in}\left(\alpha,r\right)\times\nonumber \\
 & \times & \left[1-\exp\left(-\frac{1}{2}\sigma_{\bar{q}q}\left(r\right)\int\limits _{-\infty}^{+\infty}d\zeta\rho_{A}\left(\zeta,b\right)\right)\right].\label{eq:A-frozen}\end{eqnarray}

The first term inside the square brackets in~(\ref{eq:A-frozen})
is suppressed as $\mathcal{O}(m)$, since the transition from spin-1
to spin-0 state requires helicity flip for one of the quarks in the
quark-antiquark pair, so the amplitude may be rewritten as \begin{eqnarray}
\mathcal{A}(s,\Delta_{\perp}) & \approx & -2\int d^{2}b\, e^{i\vec{\Delta}_{\perp}\cdot\vec{b}_{\perp}}\int\limits _{0}^{1}d\alpha\, d^{2}r\,\bar{\Psi}_{f}\left(\alpha,r\right)\Psi_{in}\left(\alpha,r\right)\nonumber \\
 & \times & \exp\left(-\frac{1}{2}\sigma_{\bar{q}q}\left(r\right)\int\limits _{-\infty}^{+\infty}d\zeta\rho_{A}\left(\zeta,b\right)\right)\label{eq:A-frozen-2}\end{eqnarray}

Another case when the Green function $G\left(z_{2},r_{2};z_{1},r_{1}\right)$
may be evaluated analytically is when the initial and final sizes
of dipole are small, $|r_{1}|\sim|r_{2}|\ll R_{0}(s).$ In this case
one can approximate in the rhs of Eq.~(\ref{eq:W-definition}) \begin{equation}
\sigma_{\bar{q}q}(r)\approx C\, r^{2},\label{eq:sigma-small-r}\end{equation}
 so the solution corresponds to the Green function of an oscillator
with a complex frequency,\begin{eqnarray}
 &  & G\left(z_{2},r_{2};z_{1},r_{1}\right)=\frac{a}{2\pi i\sin\left(\omega\Delta z\right)}\label{eq:W-oscillator}\\
 & \times & \exp\left(\frac{ia}{2\sin\left(\omega\Delta z\right)}\left[\left(r_{1}^{2}+r_{2}^{2}\right)\cos\left(\omega\Delta z\right)-2\vec{r}_{1}\cdot\vec{r}_{2}\right]\right),\nonumber \end{eqnarray}
 \begin{eqnarray*}
\omega^{2} & = & \frac{-2iC\rho_{A}}{\nu\alpha(1-\alpha)},\\
a^{2} & = & -iC\rho_{A}\nu\alpha(1-\alpha)/2.\end{eqnarray*}
 Then for the functions $F_{1,2}$ we can obtain explicit expressions,

\begin{widetext}\begin{eqnarray}
F_{1}\left(b,z\right) & = & \int\limits _{0}^{1}d\alpha d^{2}r_{1}d^{2}r_{2}\bar{\Psi}_{f}\left(\alpha,\vec{r}_{2}\right)\frac{a}{2\pi i\sin\left(\omega\Delta z\right)}\exp\left(\frac{ia}{2\sin\left(\omega\Delta z\right)}\left[\left(r_{1}^{2}+r_{2}^{2}\right)\cos\left(\omega\Delta z\right)-2\vec{r}_{1}\cdot\vec{r}_{2}\right]\right)_{\Delta z=z_{\infty}-z}\\
 & \times & \sigma_{\bar{q}q}\left(\vec{r}_{1},s\right)\Psi_{in}\left(\alpha,\vec{r}_{1}\right),\nonumber \\
F_{2}\left(b,z\right) & = & \int\limits _{-\infty}^{z}dz_{2}\int\limits _{0}^{1}d\alpha\, d^{2}r_{1}d^{2}r_{2}d^{2}r_{3}\bar{\Psi}_{f}\left(\alpha,\vec{r}_{3}\right)\sigma_{\bar{q}q}\left(\vec{r}_{2},s\right)\rho_{A}\left(b,z_{2}\right)\sigma_{\bar{q}q}\left(\vec{r}_{1},s\right)\label{eq:F2-incoh-res}\\
 & \times & \frac{a}{2\pi i\sin\left(\omega\Delta z\right)}\exp\left(\frac{ia}{2\sin\left(\omega\Delta z\right)}\left[\left(r_{3}^{2}+r_{2}^{2}\right)\cos\left(\omega\Delta z\right)-2\vec{r}_{3}\cdot\vec{r}_{2}\right]\right)_{\Delta z=z_{\infty}-z_{2}}\nonumber \\
 & \times & \frac{a}{2\pi i\sin\left(\omega\Delta z\right)}\exp\left(\frac{ia}{2\sin\left(\omega\Delta z\right)}\left[\left(r_{1}^{2}+r_{2}^{2}\right)\cos\left(\omega\Delta z\right)-2\vec{r}_{1}\cdot\vec{r}_{2}\right]\right)_{\Delta z=z_{2}-z}\Psi_{in}\left(\alpha,\vec{r}_{1}\right).\nonumber \end{eqnarray}
 \end{widetext}

In our evaluations of $G\left(z_{2},r_{2};z_{1},r_{1}\right)$ we
used a numerical procedure discussed in detail in~\cite{nemchik}.
We would like to emphasize that in contrast to our previous evaluation
of the pion production on a proton \cite{Kopeliovich:2011rk}, in
this paper we do not introduce by hand the coherence lengths of the
pion and effective axial meson state. They appear effectively after
convolution with proper distribution amplitudes.

In addition to the coherent processes which leaves the recoil nucleus
intact, a large contribution to pion production comes from incoherent
pion production, where the target nucleus breaks up to fragments without
particle production, like in quasi-elastic scattering. In this case
one can employ completeness of the final states, which greatly simplifies
the calculations. The analysis of such processes for the electroproduction
of vector mesons was done in~\cite{Kopeliovich:2001xj}. Its extension
to the neutrino-production is straightforward and yields \begin{equation}
\frac{d\sigma_{\nu A\to l\pi A^{*}}}{dtd\nu dQ^{2}}=\int d^{2}b\, dz\, e^{ib\cdot\Delta_{\perp}}\rho_{A}\left(b,z\right)\left|F_{1}(b,z)-F_{2}(b,z)\right|^{2}.\label{eq:XSec_incoh}\end{equation}

Differently from the coherent case, the energy dependence of the cross-section
is controlled only by the coherence length $l_{c}^{a}$, related to
the heavy axial state.

\subsection{Gluon shadowing}

\label{sec:gluon shadowing}

As was mentioned above, the presence of higher Fock components containing
gluons leads to an additional suppression caused by multiple interactions
of the gluons. This suppression is called gluon shadowing. It is controlled
by a new length scale $l_{c}^{g}$, which turns out to be much shorter
than the coherence length for quarks. As a result, no gluon shadowing
is possible at Bjorken $x>10^{-2}$. Notice that at small $Q^{2}$
Bjorken $x$ is not a proper variable, and one should switch to the
energy dependence. In this case the analog of gluon shadowing is the
Gribov inelastic shadowing correction \cite{Gribov:1984tu} related
to triple-Pomeron diffraction.

In terms of the parton model one can interpret gluon shadowing as
fusion of gluons originated from different bound nucleons in the nucleus.
Such a nonlinear effect leads to a reduction of the gluon density
at small $x$ compared with an additive density \cite{Kancheli:1973vc,mueller-qiu,Gribov:1984tu}.
While the quark distribution is directly measured in DIS, gluons can
be probed only via evolution, and this is why measurement of gluon
shadowing is still a challenge. The leading order analysis \cite{Eskola:1998df}
based on the DGLAP evolution was found to be insensitive to gluon
shadowing. Inclusion of data on hadronic collisions made the analyses
\cite{eps08,Eskola:2009uj} dependent on debatable theoretical models,
and led to such a strong gluon shadowing that the unitarity bound
\cite{bound} was severely broken. The next-to-leading order fit \cite{deFlorian:2008mr}
succeeded to constrain the gluon shadowing correction at a rather
small magnitude.

\begin{figure}
\includegraphics[scale=0.4]{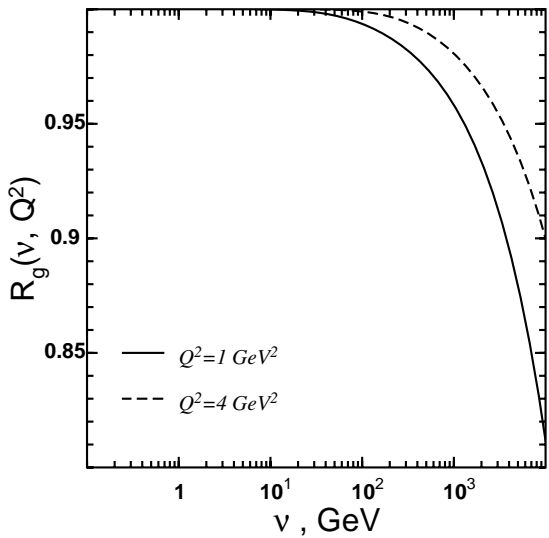}\caption{\label{fig:GluonShadowing}Gluon shadowing as a function of energy.
See~\cite{kst2,Kopeliovich:2001hf} for details of evaluation.}

\end{figure}

The theoretical predictions for gluon shadowing strongly depend on
the implemented model--while for $x\gtrsim10^{-2}$ they all predict
that the gluon shadowing is small or absent, for $x\lesssim10^{-2}$
the predictions vary in a wide range (see the review~\cite{Armesto:2006ph}
and references therein). Evaluation of the gluon shadowing within
the color dipole model was performed in~\cite{kst2,Kopeliovich:2001hf}.
In Fig.~\ref{fig:GluonShadowing} the ratio of gluon distribution
functions \[
R_{g}\left(\nu,Q^{2}\right)=\left.\frac{g_{A}\left(x,Q^{2}\right)}{A\, g_{N}\left(x,Q^{2}\right)}\right|_{x=Q^{2}/2m_{N}\nu}\]
 is plotted as function of energy $\nu$ for lead ($A=208$). We see
that in the energy range $\nu\lesssim100$~GeV gluon shadowing gives
a small correction of the order of few percent, so it can be safely
neglected.

\section{Results and discussion}

\label{sec:Results}

In this section we present the results of numerical calculations.
In this paper we make predictions in the kinematics of the ongoing
Minerva experiment at Fermilab~\cite{Drakoulakos:2004gn,McFarland:2006pz}.
While\textbf{ }there are other experiments with energies even higher
than for Minerva, they have a much worse statistics~\cite{Bell:1978qu,Allen:1985ti}.

\begin{figure*}
\includegraphics[scale=0.4]{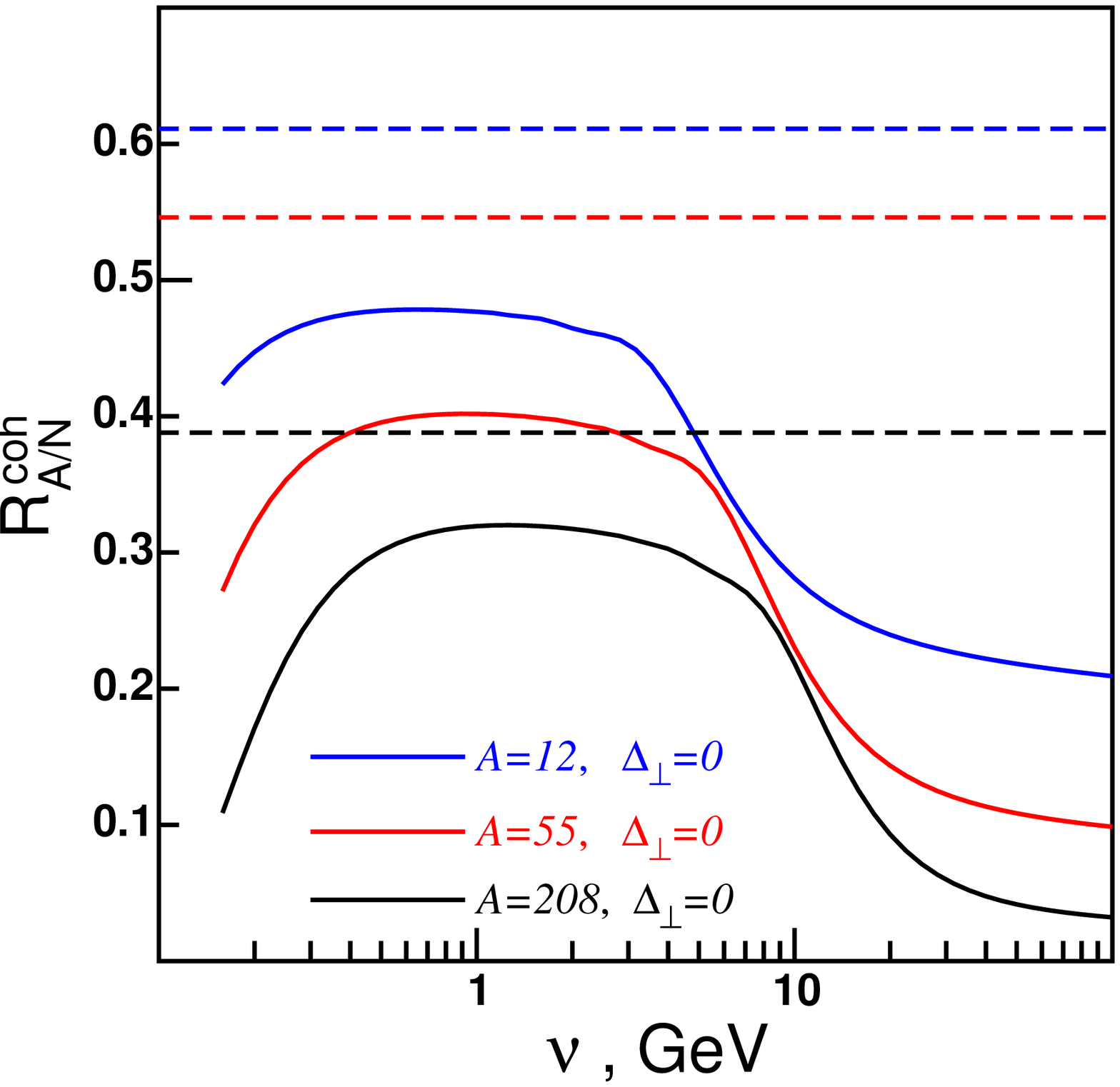}
\hspace{1cm} \includegraphics[scale=0.4]{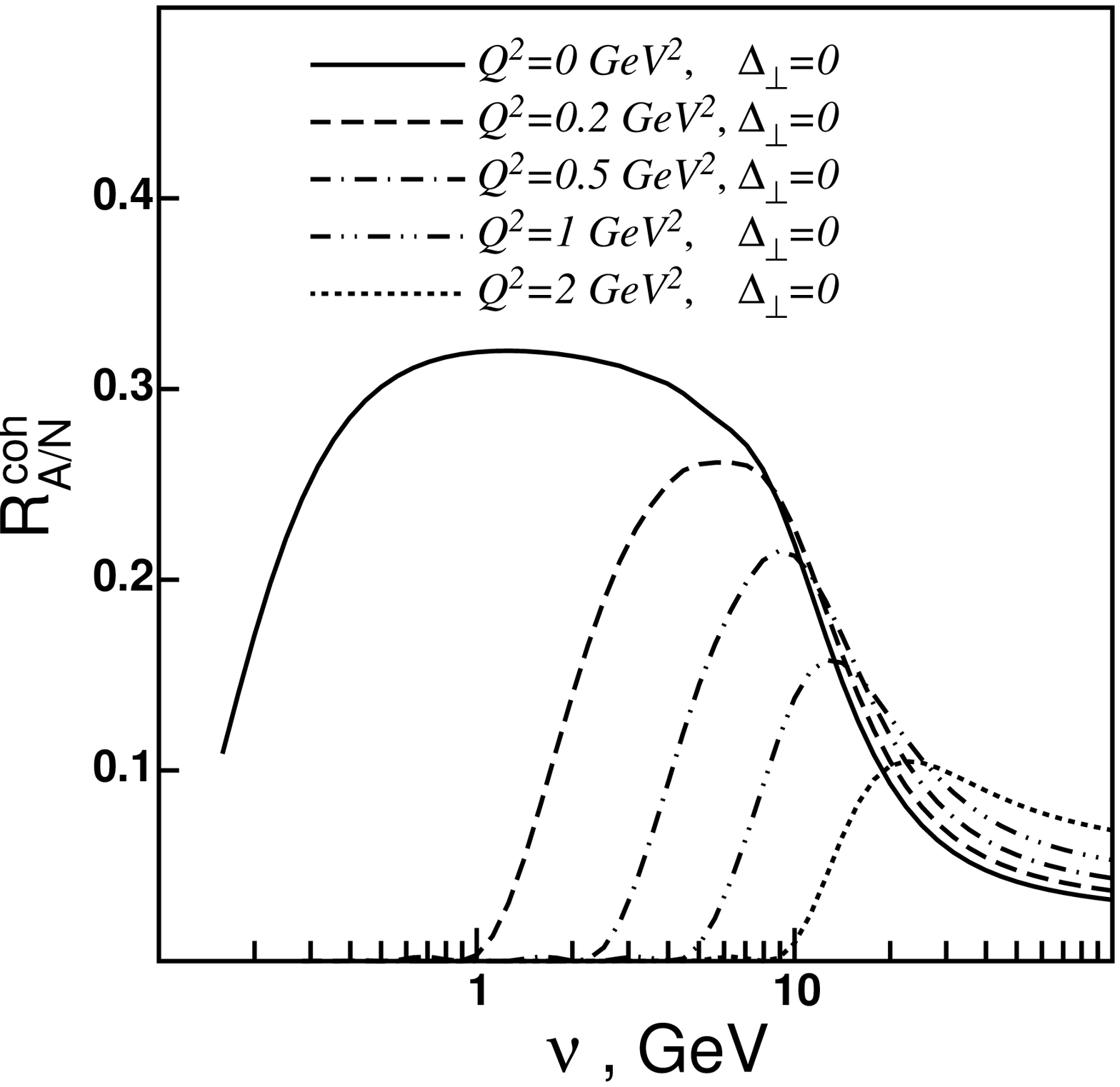}

\caption{\label{fig:W-NUCL--diff}{[}Color online{]} $\nu$-dependence of the
ratio of the coherent forward neutrino-pion production cross-sections
on nuclear and proton targets. {\sl Left:} $\nu$-dependence
of the ratio for different nuclei at $Q^{2}=0$. Solid curves correspond to the
color dipole model, dashed lines show the predictions of the Adler relation. 
{\sl Right:} $\nu$-dependence of the nuclear ratio vs  $Q^{2}$
for lead $(A=208)$.}

\end{figure*}

In Fig.~\ref{fig:W-NUCL--diff} the ratio of the cross-sections on
nuclei and nucleon, \begin{equation}
R_{A/N}^{coh}(t,\nu,Q^{2})=\frac{d\sigma_{\nu A\to l\pi A}/dtd\nu dQ^{2}}{A^{2}\, d\sigma_{\nu N\to l\pi N}/dtd\nu dQ^{2}},\label{coh}\end{equation}
 is plotted as function of transferred energy $\nu$. In the same
Figure we plotted with dashed lines predictions of the Adler relation.
As was discussed in Sect.~\ref{sec:NuclearEffects}, diffractive
pion production on nuclei is characterized by three physically distinct
energy intervals, controlled by the coherence lengths related to the
masses of pion and heavy axial states. Indeed, one can see in the
left pane, the cross-section has three different regimes. The low
energy region, $\nu\lesssim1$ GeV, is controlled by the pion coherence
length, and the cross section is suppressed if $l_{c}^{\pi}$ is short.
Notice, however, that at these low energies the dipole description
is rather formal, because one should take into account the resonances
like was done in~\cite{Lalakulich:2006sw,Lalakulich:2006yn}.

In the region $1\lesssim\nu\lesssim10$~GeV the final $\bar{q}q$
dipole (pion) is produced momentarily, and the absorptive corrections
suppress the cross section of neutrino-production of pions qualitatively
in the same way as shadowing does in the pion-nucleus elastic cross
section. However, as will be discussed below, due to the fact that
in color dipole model there is no explicit axial meson states, the
cross-section is  up to thirty percent less than the prediction given
by the Adler relation~(\ref{eq:Adler}). This issue will be discussed
in detail a few paragraphs below.

In the region $\nu\gtrsim10$~GeV all the coherence time scales become
long, so the $\bar{q}q$ dipole is produced long in advance of the
interaction and propagates through the whole nucleus. In this case
the absorptive corrections reach the maximal strength and suppress
the cross section considerably. One can see that in the right pane
of Fig.~\ref{fig:W-NUCL--diff}. The plotted results also show that
the plateau contracts when $Q^{2}$ increases, in agreement with coherence
lengths dependence on~$Q^{2}$ given in~(\ref{eq:l_pi},\ref{eq:l_a}).

\begin{figure}
\includegraphics[scale=0.4]{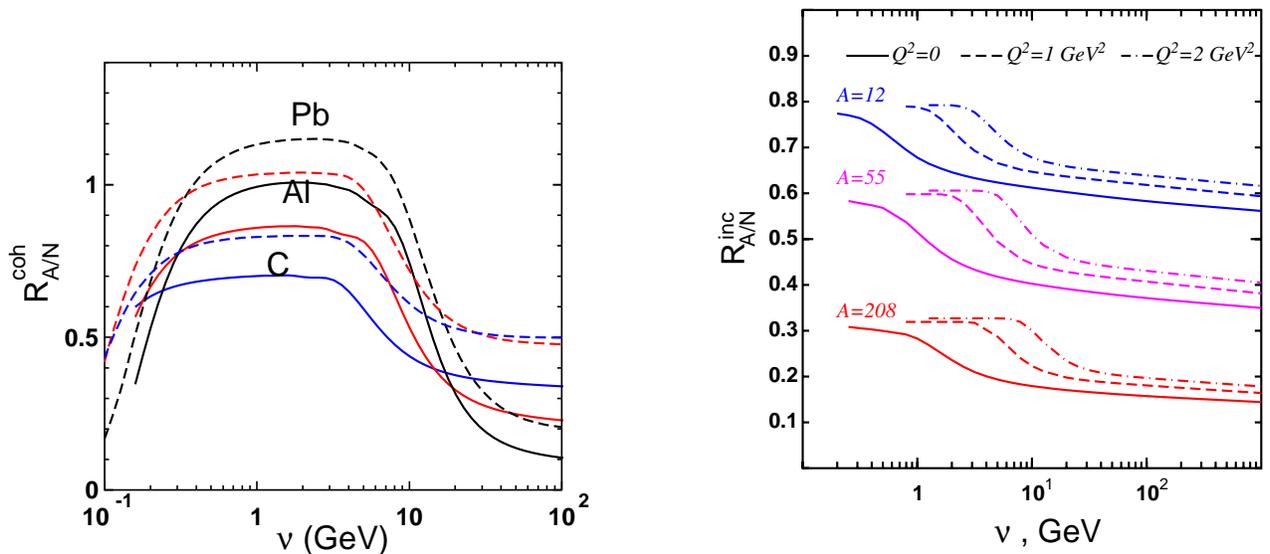}\caption{\label{fig:CDvs2Ch}{[}Color online{]} Comparison of the nuclear effects predicted 
in~\cite{Kopeliovich:2011rk} (dashed curves) and in this paper
(solid curves). }

\end{figure}

In the Figure~\ref{fig:CDvs2Ch} we compare the results for the ratio
\begin{equation}
R_{A/N}^{coh}(\nu,Q^{2})=\frac{1}{A}\frac{d^{2}\sigma_{A}/d\nu\, dQ^{2}}{d^{2}\sigma_{N}/d\nu\, dQ^{2}},\end{equation}
 plotted by solid curves vs energy $\nu$, with the expectations based
on the Adler relation shown by dashed lines. Our results are below
the Adler relation predictions at all energies. As was explained in
\cite{Kopeliovich:2011rk}, at low energy the amplitudes of pion production
on different nucleons are out of coherence, because the longitudinal
momentum transfer is large. At high energies, according to \cite{Kopeliovich:2011rk},
the lifetime of the intermediate heavy states ($a_{1}$ meson, $\rho\pi$,
etc.) is long, and absorptive corrections suppress the coherent cross
section.

There is, however, a wide energy interval from few hundreds MeV up
to about $10\GeV$, where the Adler relation was predicted to be at
work \cite{Kopeliovich:2011rk}. Now we see that even at these energies
the Adler relation is broken. To understand why this happens notice
that an effective two-channel model used in \cite{Kopeliovich:2011rk}
assumed dominance of two states in the dispersion relation for the
axial current, the pion and an effective axial vector pole $a$ with
the mass of the order of $1\GeV$. The condition of validity of the
Adler relation was shortness of the coherence length related to the
mass of the $a$-state compared to the nuclear size, \begin{equation}
l_{c}^{a}=\frac{2\nu}{Q^{2}+m_{a}^{2}}\ll R_{A}.\label{la}\end{equation}
 In contrast to the simple model, the invariant mass of a $\bar{q}q$
dipole is not fixed, $m_{\bar{q}q}^{2}=(m_{q}^{2}+k_{T}^{2})/\alpha(1-\alpha)$,
where $\alpha$ is the fractional light-cone momentum of the quark.
Correspondingly, the related coherence length, $l_{c}^{\bar{q}q}$
is distributed over a wide mass range, and while the center of the
distribution and large masses lead to a short $l_{c}^{\bar{q}q}$,
the low-mass tail of this distribution results in a long $l_{c}^{\bar{q}q}\gg R_{A}$.
For this reason the absorption corrections suppress the cross section,
as we see in Fig.~\ref{fig:CDvs2Ch}, even at moderate energies.

\begin{figure}
\includegraphics[scale=0.4]{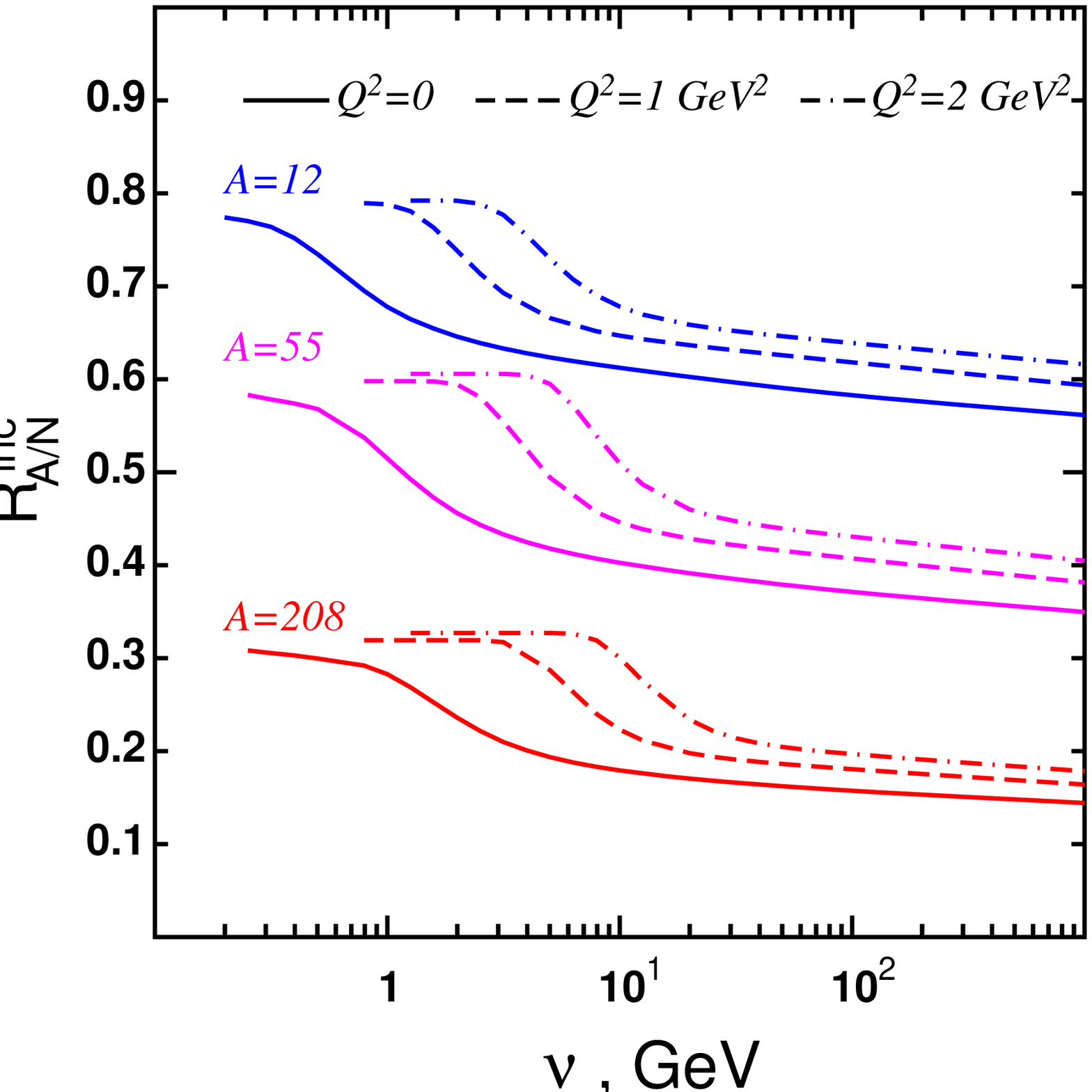}

\caption{\label{fig:W-NUCL-incoh}{[}Color online{]} $\nu$-dependence of the
ratio of the incoherent forward pion neutrino-production cross-sections 
on nuclear and proton targetsat different virtualities $Q^{2}$.}

\end{figure}

In Fig.~\ref{fig:W-NUCL-incoh} the ratio of the incoherent nuclear-to-nucleon
cross-sections \begin{equation}
R_{A/N}^{inc}(t,\nu,Q^{2})=\frac{d\sigma_{\nu A\to l\pi A^{*}}/dtd\nu dQ^{2}}{A\, d\sigma_{\nu N\to l\pi N}/dtd\nu dQ^{2}},\label{incoh}\end{equation}
 is plotted versus energy $\nu$. As was discussed in Section~\ref{sec:NuclearEffects},
the energy dependence of the incoherent cross-section is controlled
only by the shortest coherence length $l_{c}^{a}$, related to the
heavy axial states, so there are only two regimes: $l_{c}^{a}\le R_{A}$
and $l_{c}^{a}>R_{A}$. Our numerical calculations confirm such a
behavior.

Interesting that in this case of incoherent pion production the Adler
relation turns out to be severely broken at low energy, but is restored
at high energies when $l_{c}^{a}\gg R_{A}$, i.e. demonstrating a
trend opposite to coherent production. Indeed, according to the Adler
relation the cross section of incoherent pion production is proportional
to the cross section of quasi-elastic pion-nucleus scattering $\pi A\to\pi A^{*}$,
where the projectile pion must propagate and survive through the whole
nuclear thickness. The same occurs with the $\bar{q}q$ dipole in
the case of $l_{c}^{a}\gg R_{A}$.

Comparison of Figs.~\ref{fig:W-NUCL--diff} and \ref{fig:W-NUCL-incoh}
shows that in the forward kinematics ($\Delta_{\perp}=0$) the coherent
cross-section is much higher than the incoherent one. However, for
the off-forward case the coherent cross-section is suppressed by the
nuclear formfactor, whereas the incoherent cross-section is controlled
by the proton formfactor. For this reason, for sufficiently large
values of $t=\Delta^{2}$ the incoherent cross-section surpasses the
coherent one.

\section{Summary}

We performed calculations for the nuclear effects in diffractive neutrino-production
of pions basing on the color-dipole description. The non-zero phase
shifts between the production amplitudes on different bound nucleons
are taken into account applying the path integral technique. The results
confirmed the presence of prominent structures in the energy dependence
of coherent and incoherent processes on nuclei. Although the general
pattern of nuclear effects agrees with what was predicted in \cite{Kopeliovich:2011rk}
within a simple 2-channel model, the new important features are observed
basing on the more detailed dynamics of the dipole description. Namely,
while the effective 2-channel model predicted validity of the Adler
relation in the wide energy interval from few hundreds MeV up to about
$10\GeV$ \cite{Kopeliovich:2011rk}, with the dipole approach we
found a considerable suppression of the nuclear cross section compared
to the result of the Adler relation within the same energy interval.
This happens, due to a contribution of light dipoles possessing a
longer coherence length compared to the fixed mass heavy intermediate
state assumed in the 2-channel model. The contribution of such light
dipoles is subject to strong absorptive corrections reducing the cross
section. At the same time, at high energies both models predict a
similar strong breakdown of the Adler relation.
\begin{acknowledgments}
This work was supported in part by Fondecyt (Chile) grants 1090291,
1100287 and 1120920, and by Conicyt-DFG grant No. 084-2009. \end{acknowledgments}


\begin{thebibliography}{70}
\bibitem{Drakoulakos:2004gn}D.~Drakoulakos \emph{et al.} {[}Minerva
Collaboration{]}, 
 arXiv:hep-ex/0405002.

\bibitem{Adler:1964yx}S.~L.~Adler, 
 Phys.~Rev.~ \textbf{135} (1964) B963.

\bibitem{Adler:1966gc}S.~L.~Adler and Y.~Dothan, 
 Phys.~Rev.~\textbf{151} (1966) 1267.

\bibitem{belkov} A.~A.~Belkov and B.~Z.~Kopeliovich, 
 Sov.\ J.\ Nucl.\ Phys.\ \textbf{46}, 499 (1987) {[}Yad.\ Fiz.\ \textbf{46},
874 (1987){]}.

\bibitem{Rein:1982pf}D.~Rein and L.~M.~Sehgal, 
 Nucl.~Phys.~B \textbf{223} (1983) 29.

\bibitem{Kopeliovich:2011rk}B.~Z.~Kopeliovich, I. Potashnikova,
M.~Siddikov, I.~Schmidt, Phys.~Rev.~C \textbf{84} (2011) 024608
{[}arXiv:1105.1711 {[}hep-ph{]}{]}.

\bibitem{bell} J. Bell, Phys. Rev. Lett. \textbf{13} 57 (1964).

\bibitem{glauber} R. J. Glauber, in \textit{Lectures in Theoretical
Physics}, W. E. Brittin \textit{et al} Editors, New York (1959).

\bibitem{Gribov:1968jf}V.~N.~Gribov, 
 Sov.~Phys.~JETP \textbf{29} (1969) 483 {[}Zh.~Eksp.~Teor.~Fiz.~\textbf{56}
(1969) 892{]}.

\bibitem{k-shad} B.~Z.~Kopeliovich, 
 Phys.\ Lett.\ B \textbf{227}, 461 (1989).

\bibitem{gransasso} B.~Z.~Kopeliovich, 
 Nucl.\ Phys.\ Proc.\ Suppl.\ \textbf{139}, 219 (2005).

\bibitem{shad-jetp} B.~Z.~Kopeliovich, 
 Sov.\ Phys.\ JETP \textbf{70} (1990) 801 {[}Zh.\ Eksp.\ Teor.\ Fiz.\ \textbf{97}
(1990) 1418{]}.

\bibitem{wa59} WA59 Collaboration, P.P. Allport et al., Phys. Lett.
B \textbf{232} 417 (1989).

\bibitem{Kopeliovich:1981pz}B.~Z.~Kopeliovich, L.~I.~Lapidus
and A.~B.~Zamolodchikov, JETP Lett. \textbf{33}, 595 (1981) {[}Pisma
Zh. Eksp. Teor. Fiz. \textbf{33}, 612 (1981){]}.

\bibitem{Lalakulich:2006sw}O.~Lalakulich, E.~A.~Paschos and G.~Piranishvili,
Phys.~Rev.~D \textbf{74} (2006) 014009 {[}arXiv:hep-ph/0602210{]}.

\bibitem{Lalakulich:2006yn}O.~Lalakulich, W.~Melnitchouk and E.~A.~Paschos,
Phys.~Rev.~C \textbf{75} (2007) 015202 {[}arXiv:hep-ph/0608058{]}.

\bibitem{Nakamura:2011rt}S.~X.~Nakamura, arXiv:1109.4443 {[}nucl-th{]}.

\bibitem{GolecBiernat:1998js} K.~J.~Golec-Biernat and M.~W\"usthoff,
Phys.\ Rev.\ D \textbf{59} (1999) 014017 {[}arXiv:hep-ph/9807513{]}.

\bibitem{GolecBiernat:2004xu}K.~J.~Golec-Biernat, Acta Phys.\ Polon.\ B
\textbf{35}, 3103 (2004).

\bibitem{raufeisen} J.~Raufeisen, J.~-C.~Peng, G.~C.~Nayak,
Phys.\ Rev.\ \textbf{D66}, 034024 (2002). {[}hep-ph/0204095{]}.

\bibitem{Hufner:2000jb}J.~Hufner, Yu.~P.~Ivanov, B.~Z.~Kopeliovich
and A.~V.~Tarasov,Phys. Rev. D \textbf{62} (2000) 094022 {[}arXiv:hep-ph/0007111{]}.

\bibitem{Kopeliovich:2008ct}B.~Z.~Kopeliovich, I.~Schmidt and
M.~Siddikov, Phys.~Rev.~D \textbf{79} (2009) 034019 {[}arXiv:0812.3992
{[}hep-ph{]}{]}.

\bibitem{Kopeliovich:2009cx}B.~Z.~Kopeliovich, I.~Schmidt and
M.~Siddikov, Phys.~Rev.~D \textbf{80} (2009) 054005 {[}arXiv:0906.5589
{[}hep-ph{]}{]}.

\bibitem{Kopeliovich:2010xm}B.~Z.~Kopeliovich, I.~Schmidt and
M.~Siddikov, 
 Phys.~Rev.~D \textbf{82} (2010) 014017 {[}arXiv:1005.4621 {[}hep-ph{]}{]}.

\bibitem{Kopeliovich:2010sa}B.~Z.~Kopeliovich, I.~Schmidt and
M.~Siddikov, 
 Phys.~Rev.~D \textbf{81} (2010) 094013 {[}arXiv:1003.4188 {[}hep-ph{]}{]}.

\bibitem{Machado:2008tp}M.~V.~T.~Machado, 
 Eur.~Phys.~J.~C \textbf{59} (2009) 769 {[}arXiv:0810.3665 {[}hep-ph{]}{]}.

\bibitem{Machado:2008zv}M.~V.~T.~Machado, 
 Phys.~Rev.~D~\textbf{78} (2008) 034016 {[}arXiv:0805.3144 {[}hep-ph{]}{]}.

\bibitem{Machado:2009cd}M.~V.~T.~Machado, 
 arXiv:0905.4516 {[}hep-ph{]}.

\bibitem{Fiore:2005bp}R.~Fiore and V.~R.~Zoller, Phys.~Lett.~B
\textbf{632} (2006) 87 {[}arXiv:hep-ph/0509097{]}.

\bibitem{Fiore:2005yi}R.~Fiore and V.~R.~Zoller, JETP Lett.~\textbf{82}
(2005) 385 {[}Pisma Zh.~Eksp.~Teor.~Fiz.~\textbf{82} (2005) 440{]}
{[}arXiv:hep-ph/0508187{]}.

\bibitem{Fiore:2008cc}R.~Fiore and V.~R.~Zoller, JETP Lett.~\textbf{87}
(2008) 524 {[}arXiv:0803.4492 {[}hep-ph{]}{]}.

\bibitem{Fiore:2008nj}R.~Fiore and V.~R.~Zoller, Phys.~Lett.~B
\textbf{681} (2009) 32 {[}arXiv:0812.4501 {[}hep-ph{]}{]}.

\bibitem{GayDucati:2008hi}M.~B.~Gay Ducati, M.~M.~Machado and
M.~V.~T.~Machado, Phys.~Rev.~D \textbf{79} (2009) 073008 {[}arXiv:0812.4273
{[}hep-ph{]}{]}.

\bibitem{GayDucati:2008zzc}M.~B.~Gay Ducati, M.~M.~Machado and
M.~V.~T.~Machado, Braz.~J.~Phys.~\textbf{38} (2008) 487.

\bibitem{Ducati:2006vh}M.~B.~G.~Ducati, M.~M.~Machado and M.~V.~T.~Machado,
Phys.~Lett.~B \textbf{644} (2007) 340 {[}arXiv:hep-ph/0609088{]}.

\bibitem{Machado:2007wq}M.~V.~T.~Machado, Phys.~Rev.~D \textbf{75}
(2007) 093008 {[}arXiv:hep-ph/0703111{]}.

\bibitem{Kopeliovich:2011xw}B.~Z.~Kopeliovich, I.~Schmidt and
M.~Siddikov, Phys.~Rev.~D \textbf{84} (2011) 033012 {[}arXiv:1107.2845
{[}hep-ph{]}{]}.

\bibitem{Dorokhov:2006qm} A.~E.~Dorokhov, W.~Broniowski and E.~Ruiz
Arriola, 
 Phys.\ Rev.\ D \textbf{74} (2006) 054023 {[}arXiv:hep-ph/0607171{]}.

\bibitem{Anikin:2000rq}I.~V.~Anikin, A.~E.~Dorokhov and L.~Tomio,
Phys.~Part.~Nucl. \textbf{31} (2000) 509 {[}Fiz.~Elem.~Chast.~Atom.~Yadra
\textbf{31} (2000) 1023{]}.

\bibitem{Dorokhov:2003kf}A.~E.~Dorokhov and W.~Broniowski, Eur.~Phys.~J.~C
\textbf{32} (2003) 79 {[}arXiv:hep-ph/0305037{]}.

\bibitem{Goeke:2007j}K.~Goeke, M.~M.~Musakhanov and M.~Siddikov,
Phys.~Rev.~D \textbf{76} (2007) 076007 {[}arXiv:0707.1997 {[}hep-ph{]}{]}

\bibitem{Kopeliovich:2011rv}B.~Z.~Kopeliovich, I.~Schmidt and
M.~Siddikov, arXiv:1108.5654 {[}hep-ph{]}.

\bibitem{Buttimore:1998rj}N.~H.~Buttimore, B.~Z.~Kopeliovich,
E.~Leader, J.~Soffer and T.~L.~Trueman, Phys.~Rev.~D \textbf{59}
(1999) 114010 {[}arXiv:hep-ph/9901339{]}.

\bibitem{Kopeliovich:2007fv} B.~Z.~Kopeliovich, H.~J.~Pirner,
A.~H.~Rezaeian and I.~Schmidt, 
 Phys.\ Rev.\ D \textbf{77} (2008) 034011 {[}arXiv:0711.3010 {[}hep-ph{]}{]}.

\bibitem{Kopeliovich:2008nx} B.~Z.~Kopeliovich, A.~H.~Rezaeian
and I.~Schmidt, 
 Phys.~Rev.~D \textbf{78} (2008) 114009 {[}arXiv:0809.4327 {[}hep-ph{]}{]}.

\bibitem{Kopeliovich:2008da}B.~Z.~Kopeliovich, I.~K.~Potashnikova,
I.~Schmidt and J.~Soffer, 
 Phys.~Rev.~D~\textbf{78} (2008) 014031 {[}arXiv:0805.4534 {[}hep-ph{]}{]}.

\bibitem{Bronzan:1974jh}J.~B.~Bronzan, G.~L.~Kane and U.~P.~Sukhatme,
Phys.~Lett.~B \textbf{49} (1974) 272.

\bibitem{bjorken} J.~D.~Bjorken and J.~B.~Kogut, 
 Phys.\ Rev.\ D \textbf{8}, 1341 (1973).

\bibitem{k-povh} B.~Kopeliovich and B.~Povh, Phys.\ Lett.\ B
\textbf{367}, 329 (1996); Z.\ Phys.\ A \textbf{356}, 467 (1997)
{[}arXiv:nucl-th/9607035{]}.

\bibitem{asymmetry} B.~Z.~Kopeliovich, I.~K.~Potashnikova, I.~Schmidt
and J.~Soffer, 
 arXiv:1109.2500 {[}hep-ph{]}; to appear in Phys. Rev. D.

\bibitem{deck} R.~T.~Deck, 
 Phys.\ Rev.\ Lett.\ \textbf{13}, 169 (1964).

\bibitem{marage} B.~Z.~Kopeliovich and P.~Marage, 
 Int.\ J.\ Mod.\ Phys.\ A \textbf{8}, 1513 (1993).

\bibitem{Kopeliovich:1998gv}B.~Z.~Kopeliovich, J.~Raufeisen and
A.~V.~Tarasov, 
 \textbf{440} (1998) 151 {[}arXiv:hep-ph/9807211{]}.

\bibitem{kz91} B.~Z.~Kopeliovich and B.~G.~Zakharov, 
 Phys.\ Rev.\ D \textbf{44}, 3466 (1991).

\bibitem{kst2} B.~Z.~Kopeliovich, A.~Sch\"afer and A.~V.~Tarasov,
Phys.\ Rev.\ D \textbf{62}, 054022 (2000).

\bibitem{nemchik} J.~Nemchik, 
 Phys.\ Rev.\ C \textbf{68}, 035206 (2003) {[}arXiv:hep-ph/0301043{]}.

\bibitem{Kopeliovich:2001xj}B.~Z.~Kopeliovich, J.~Nemchik, A.~Sch\"afer
and A.~V.~Tarasov, Phys.~Rev.~C \textbf{65} (2002) 035201 {[}arXiv:hep-ph/0107227{]}.

\bibitem{Gribov:1984tu} L.~V.~Gribov, E.~M.~Levin and M.~G.~Ryskin,
Phys.\ Rept.\ \textbf{100} (1983) 1.

\bibitem{Kancheli:1973vc}O.~V.~Kancheli, Pisma Zh.~Eksp.~Teor.~Fiz.~\textbf{18}
(1973) 465.

\bibitem{mueller-qiu} A.~H.~Mueller and J.~-w.~Qiu, 
 Nucl.\ Phys.\ B\ \textbf{268}, 427 (1986).

\bibitem{Eskola:1998df}K.~J.~Eskola, V.~J.~Kolhinen and C.~A.~Salgado,
Eur.~Phys.~J.~C \textbf{9} (1999) 61 {[}arXiv:hep-ph/9807297{]}.

\bibitem{eps08}K.~J.~Eskola, H.~Paukkunen and C.~A.~Salgado,
JHEP \textbf{0807}, 102 (2008) {[}arXiv:0802.0139 {[}hep-ph{]}{]}.

\bibitem{Eskola:2009uj}K.~J.~Eskola, H.~Paukkunen and C.~A.~Salgado,
JHEP \textbf{0904} (2009) 065 {[}arXiv:0902.4154 {[}hep-ph{]}{]}.

\bibitem{bound}B.~Z.~Kopeliovich, E.~Levin, I.~K.~Potashnikova
and I.~Schmidt, 
 \textbf{79}, 064906 (2009) {[}arXiv:0811.2210 {[}hep-ph{]}{]}.

\bibitem{deFlorian:2008mr}D.~de Florian, R.~Sassot, M.~Stratmann
and W.~Vogelsang, Phys.~Rev.~Lett.~\textbf{101} (2008) 072001
{[}arXiv:0804.0422 {[}hep-ph{]}{]}.

\bibitem{Armesto:2006ph}N.~Armesto, J.~Phys.~G\textbf{ 32} (2006)
R367 {[}arXiv:hep-ph/0604108{]}.

\bibitem{Kopeliovich:2001hf}B.~Z.~Kopeliovich, J.~Raufeisen, A.~V.~Tarasov
and M.~B.~Johnson, Phys.~Rev.~C \textbf{67} (2003) 014903 {[}arXiv:hep-ph/0110221{]}.

\bibitem{McFarland:2006pz}K.~S.~McFarland {[}MINERvA Collaboration{]},
Nucl.~Phys.~Proc.~Suppl.~\textbf{159} (2006) 107 {[}arXiv:physics/0605088{]}.

\bibitem{Bell:1978qu}J.~Bell \emph{et al.}, Phys.~Rev.~Lett.~\textbf{41}
(1978) 1008.

\bibitem{Allen:1985ti}P.~Allen \emph{et al.} {[}Aachen-Birmingham-Bonn-CERN-London-Munich-Oxford
Collaboration{]}, Nucl.~Phys.~B \textbf{264} (1986) 221. 
\end{thebibliography}
\end{document}